\documentclass[11pt]{article}
\usepackage{latexsym}
\begin{document}
\newcommand\lsim{\roughly{<}}
\newcommand\gsim{\roughly{>}}
\newcommand\CL{{\cal L}}
\newcommand\CO{{\cal O}}
\newcommand\half{\frac{1}{2}}
\newcommand\beq{\begin{eqnarray}}
\newcommand\eeq{\end{eqnarray}}
\newcommand\eqn[1]{\label{eq:#1}}
\newcommand\intg{\int\,\sqrt{-g}\,}
\newcommand\eq[1]{eq. (\ref{eq:#1})}
\newcommand\meN[1]{\langle N \vert #1 \vert N \rangle}
\newcommand\meNi[1]{\langle N_i \vert #1 \vert N_i \rangle}
\newcommand\mep[1]{\langle p \vert #1 \vert p \rangle}
\newcommand\men[1]{\langle n \vert #1 \vert n \rangle}
\newcommand\mea[1]{\langle A \vert #1 \vert A \rangle}
\newcommand\bi{\begin{itemize}}
\newcommand\ei{\end{itemize}}
\newcommand\be{\begin{equation}}
\newcommand\ee{\end{equation}}
\newcommand\bea{\begin{eqnarray}}
\newcommand\eea{\end{eqnarray}}
\def\Dsl{\,\raise.15ex \hbox{/}\mkern-12.8mu D}
\newcommand\Tr{{\rm Tr\,}}
\thispagestyle{empty}
\begin{titlepage}
\begin{flushright}
CALT-68-2405\\
\end{flushright}
\vspace{1.0cm}
\begin{center}
{\LARGE \bf  Implications of Correlated Default For Portfolio Allocation to Corporate Bonds\footnote{To appear in, Correlated Default Analysis for Collateralized Debt Obligations, edited by S. Das, G. Fong and N. Kapadia.}  }\\ 
\bigskip
\bigskip\bigskip
{ Mark B. Wise$^a$ and Vineer Bhansali$^b$} \\
~\\
\noindent
{\it\ignorespaces
          (a) California Institute of Technology, Pasadena CA 91125\\

          {\tt wise@theory.caltech.edu}\\
\bigskip   (b) PIMCO, 840 Newport Center Drive, Suite 300\\
               Newport Beach, CA 92660 \\

{\tt   bhansali@pimco.com}
}\bigskip
\end{center}
\vspace{1cm}
\begin{abstract}
This article deals with the problem of optimal allocation of capital to corporate bonds in fixed income portfolios when there is the possibility of correlated defaults. Using a multivariate normal Copula function for the joint default probabilities we show that retaining the first few moments of the portfolio default loss distribution gives an extremely good approximation to the full solution of the asset allocation problem. We provide detailed results on the convergence of the moment expansion and explore how the optimal portfolio allocation depends on recovery fractions, level of diversification and investment time horizon. Numerous numerical illustrations exhibit the results for simple portfolios and utility functions.
\end{abstract}
\vfill
\end{titlepage}

\section{Introduction}
Investors routinely look to the corporate bond market in particular, and spread markets in general, to enhance the performance of their portfolios. However, for every source of excess return over the risk-free rate, there is a source of excess risk. When sources of risk are correlated, the allocation decision to the risky sectors, as well as allocation to particular securities in that sector, can be substantially different from the uncorrelated case. Since the joint probability distribution of returns of a set of defaultable bonds varies with the joint probabilities of default, recovery fraction for each bond, and the number of defaultable bonds, a direct approach to the allocation problem that incorporates all these factors completely can only be attempted numerically within the context of a default model. This approach can have the short-coming of hiding the intuition behind the asset allocation process in practice, which leans very heavily on the quantification of the first few moments, such as the mean, variance and skewness. In this paper, we will take the practical approach of characterizing the portfolio default loss distribution in terms of its moment expansion. Focusing on allocation to corporate bonds (the analysis in this paper can be generalized to any risky sector which has securities with discrete payoffs), we will answer the following questions:
\begin{itemize}
\item In the presence of correlated defaults, how well does retaining the first few moments of the portfolio default loss distribution do, for the portfolio allocation problem, as compared to the more intensive full numerical solution?
\item For different choices of correlations, probabilities of default, number of bonds in the portfolio, and investment time horizon, how does the optimal allocation to the risky bonds vary?
\end{itemize}

In this paper we consider portfolios consisting of risk free assets at a return $y$ and corporate zero coupon bonds with  return $c_i$ for firm $i$, and study how correlations between defaults affect the optimal allocation to corporate assets in the portfolio over some time horizon $T$. We assume that if company $i$ defaults at time $t$ a  fraction $R_i$ (the recovery fraction) of the value of its bonds is recovered and reinvested at the risk free rate. 
We will quantify the impact of risk associated with losses from corporate defaults  \footnote{In this paper, we use annualized units so quoted default probabilities, hazard rates and corporate rates are annualized ones.} on portfolio allocation. To compensate the investor for default risk, the corporate bond's return $c_i$ is greater than that of the risk-free assets in the portfolio. The excess return, or spread of investment in bonds of firm $i$ can be decomposed into two parts. The first part of the spread, which we call $\lambda_i$,  arises 
as the actuarially fair value of assuming the risk of default. The second part, which we call $\mu_i$, is the excess risk premium that compensates the corporate bond investor above and beyond the probabilistically fair value. In practice, $\mu_i$ can arise due to a number of features not directly related to defaults.  For example, traders will partition $\mu_i$ into two pieces, one for liquidity and one for event risk, $\mu_i = l_i+e_i$.  Most low grade bonds may have a full percentage point arising simply from liquidity premia, and $\mu_i$ can fluctuate to large values in periods of credit stress. Liquidity $l_i$ is systematic and one would expect it to be roughly equal for a similar class of bonds. Event risk $e_i$ contributes to $\mu_i$ due to the firm's specific vulnerability to factors that affect it (e.g. negative press). In periods of stress the default probabilities, $l_i$ and $e_i$ all increase simultaneously, and the recovery rate expectations $R_i$ fall, leading to a spike in the overall spread.  Since these variables can be highly volatile, the reason behind the portfolio approach to managing credit is to minimize the impact of non-systematic event risk in the portfolio.\footnote{The authors would like to thank David Hinman of PIMCO for enlightening discussions on this topic.}  The excess risk premium itself is not very stable over time.  Empirical research shows that the excess risk premium might vary from tens of basis points to hundreds of basis points.  For instance, in the BB asset class, if we assume a recovery rate of 50\% and default probability of 2\%, the actuarially fair value of the spread is 100 basis points.  However, it is not uncommon to find actual spreads of the BB class to be 300 bp over treasuries [Altman (1989)].  The excess 200 bp of risk premium can be decomposed in any combination of liquidity premium and event risk premium, and is best left to the judgment of the market.  When the liquidity premium component is small compared to event risk premium, we would expect that portfolio diversification and the methods of this paper are valuable. 

 One other factor needs to be kept in mind when comparing historical spreads to current levels.  In the late eighties and early nineties, the spread was routinely quoted in terms of a treasury benchmark curve.  However, the market itself has developed to a point where spreads are quoted over both the libor swap rate and the treasury rate, and the swap rate has gradually substituted the treasury rate as the risk-free benchmark curve.  This has two impacts.  Firstly, since the swap spread (swap rate minus treasury rate for a given maturity) in US is significant (of the order of 50 bp as of this writing), the excess spread needs to be computed as a difference to the swap yield curve.  Secondly, the swap-spread itself has been very volatile during the last few years, which leads to an added source of non default related risk in the spread of corporate bonds when computed against the treasury curve.  
Thus, the 200 bp of residual spread is effectively 150 bp over the swap rate in the BB example, of which, for lack of better knowledge, equal amounts may be assumed to arise from liquidity and event risk premium over the long term.  The allocation decision to risky bonds strongly depends on the level of risk-aversion in the investor's utility function, and the required spread for a given allocation will go up nonlinearly as risk aversion increases.

The value of the optimal fraction of the portfolio in corporate bonds, here called $\alpha_{opt}$, cannot be determined without knowing the excess risk premium part of the corporate returns. They provide the incentive for a risk averse investor to choose corporate securities over risk free assets. In this paper we explore, using utility functions with constant relative risk aversion, the convergence of the moment expansion for $\alpha_{opt}$. Our work indicates that, (for $\alpha_{opt}$ less than unity)  $\alpha_{opt}$ is usually determined by the mean, variance and skewness of the portfolio default loss probability distribution. The sensitivity to higher moments increases as $\alpha_{opt}$ does. Some measures of default risk, for example a VAR analysis\footnote{See for example, Jorion (2001).}, may be more sensitive to the tail of the default loss distribution. We also examine how the optimal portfolio allocation scales with the number of firms, time horizon and recovery fractions.

Historical evidence suggests that on average default correlations increase with the time horizon\footnote{ Zhou (2001) derives an analytic formula for default correlations in a first passage-time default model and finds a similar increase with time horizon.}. For example, Lucas (1995) estimates that over one year, two year and five year time horizons default correlations between $Ba$ rated firms are $2\%$, $6\%$ and $15\%$ respectively. However, the errors in extracting default correlations from historical data are likely to be large since defaults are rare. Also these historical analysis neglect firm specific effects that may be very important for portfolios weighted towards a particular economic sector. Furthermore, in periods of market stress default probabilities and their correlations increase [Das, Freed, Geng and Kapadia (2001)] dramatically.

There are other sources of risk associated with corporate securities. For example, the market's perception of firm $i$'s probability of default could increase over the time horizon $T$ resulting in a reduction in the value of its bonds. For a recent discussion on portfolio risk due to downgrade fluctuations see Dynkin, Hyman and Konstantinovsky (2002). Here we do not address the issue of risk associated with fluctuations in the credit spread but rather focus on the risk associated with losses from actual defaults.

In the next section a simple model for default is introduced. The model assumes a multivariate normal Copula function for the joint default probabilities. In section $3$ we set up the portfolio problem. Moments of the fractional corporate default loss probability distribution are expressed in terms of joint default probabilities and it is shown how these can be used to determine $\alpha_{opt}$. In section $4$ the impact of correlations on the portfolio allocation problem is studied using sample portfolios where all the firms have the same probabilities of default and the correlations between firms are all the same. The recovery fractions are assumed to be zero for the portfolios in section $4$. The impact of non-zero recovery fractions on the convergence of the moment expansion is studied in section $5$. Concluding remarks are given in section $6$. 

This work is based on Wise and Bhansali (2002). It extends the results presented in that paper to arbitrary time horizons (allowing default to occur at any time) and makes more realistic assumptions for the consequences of default.

\section{A Model For Default}

It is convenient for discussions of default risk to introduce the random variables $\hat n_i (t_i)$. $\hat n_i (t_i)$  takes the value $1$ if firm $i$ defaults in the time horizon $ t_i$ and zero otherwise. The joint default probabilities are expectations of products of these random variables, 
\be
\label{two}
P_{i_1 \ldots i_m}(t_{i_1}, \ldots ,t_{i_m})=E[\hat n_{i_1}(t_{i_1})\cdots \hat n_{i_m}(t_{i_m})], ~~{\rm when}~~ i_1 \ne i_2 \cdots \ne i_m. 
\ee
$P_i(t_i)$ is the probability that firm $i$ defaults in the time period $t_i$ and $P_{i_1, \ldots ,i_m}(t_{i_1}, \ldots ,t_{i_m})$ is the joint probability that the $m$-firms $i_1, \dots i_m$ default in the times periods $t_{i_1}, \dots ,t_{i_m}$. 

We assume that the joint default probabilities are given by a multivariate normal Copula function [See for example, Lee (2000)]. Explicitly,
\begin{equation}
\label{key}
P_{1 \ldots n}(t_{1}, \ldots ,t_{n}) = \frac{1}{(2\pi)^{n \over 2}
\sqrt{\det \xi}}\int_{-\infty}^{-\chi_1(t_1)} dx_1 \cdots \int_{-\infty}^{-\chi_n(t_n)} dx_n \exp \left[-\frac{1}{2}\sum_{ij} {x_i \xi^{(-1)}_{ij} x_j}\right],
\end{equation}
where the sum goes over $i,j=1, \ldots ,n$, and $\xi^{(-1)}_{ij}$ is the inverse of the $n \times n$ correlation matrix $\xi_{ij}$. For $n$ not too large the integrals in equation (\ref{key}) can be done numerically or, since defaults are rare, analytic results can be obtained using the leading terms in an asymptotic expansion of the integrals. The choice of a multivariate normal Copula function is common but somewhat arbitrary. In principle the Copula function should be chosen based on a comparison with data. For recent work along these lines see Das and Geng (2002).

In equation (\ref{key}) the $n \times n$ correlation matrix $\xi_{ij}$ is usually taken to be the asset correlation matrix. This has the advantage of allowing a connection to stock prices [Merton (1974)].  However this assumption is not necessary. For example the assets, $\hat a_i$ could be functions of normal random variables, $\hat a_i=g_i(\hat z_i)$. Suppose default occurs if the assets $\hat a_i$ cross the thresholds $T_i$. Then the condition for default on the normal production factor variables $\hat z_i$ is that they cross the thresholds $g^{(-1)}_i(T_i)$. In such a model it is natural to interpret $\xi_{ij}$ as the correlation matrix for the normal production factor variables $\hat z_i$. If the functions $g_i$ are linear then the assets are also normal and their correlation matrix is also given by $\xi_{ij}$. However if the functions $g_i$ are not linear the joint probability distribution for the assets is not multivariate normal and can have fat tails. Unless the $g_i$ are specified it is not possible to connect stock prices to the correlation matrix $\xi_{ij}$ and the default thresholds $\chi_i$.  However, even when the functions $g_i$ are not known the default thresholds or ``equivalent distances to default"  $\chi_i(t_i)$ and the correlation matrix $\xi_{ij}$ are determined by the default probabilities $P_i(t_i)$ and the default correlations $d_{ij}(t)$ and so have a direct connection to measures that investors use in quantifying security risk. In our work we assume that the correlation matrix is time independent.

Equation (\ref{key}) in the case $n=1$ gives,
\begin{equation}
P_i(t_i)={1 \over (2\pi)^{1 \over 2}}\int_{-\infty}^{-\chi_i(t_i)}dx_i\exp \left[-\frac{1}{2}x_i^2\right].
\end{equation} 
Hence for an explicit choice for the time dependence of the default probabilities the default thresholds are known. 
A simple (and frequently used) choice for the default probabilities is 
\begin{equation}
\label{hazard}
P_i(t_i)=1-\exp(-h_it_i),
\end{equation}
where the hazard rates $h_i$ are independent of time. 

Since $\hat n_i(t_i)^2=\hat n_i(t_i)$ it follows that the correlation of defaults between two different firms (which we choose to label $1$ and $2$) is, 
\begin{eqnarray}
d_{12}(t)&=&{E[\hat n_1(t) \hat n_2(t)]-E[\hat n_1(t)]E[ \hat n_2(t)]\over \sqrt{(E[\hat n_1(t)^2]-E[\hat n_1(t)]^2)(E[\hat n_2(t)^2]-E[n_2(t)]^2)}} \nonumber \\
&=&{ P_{12}(t,t)-P_1(t)P_2(t)\over \sqrt{P_1(t)(1-P_1(t))P_2(t)(1-P_2(t))}}.
\end{eqnarray}

The default model we are adopting is not as well motivated as a first passage-time default model [Black and Cox (1976), Longstaff and Schwartz (1995), Leland and Toft (1996), {\it etc}.] where the assets undergo a random walk and default is associated with the first time that the assets fall below the liabilities. However, it is very convenient to work with.

We will evaluate the integrals in equation (\ref{key}) by numerical integration. For this we need the inverse and determinant of the correlation matrix. It is very important that the correlation matrix $\xi_{ij}$ is positive semi-definite. If it has negative eigenvalues the integrals in equations (\ref{key}) are not well defined. Typically a correlation matrix that is forecast using qualitative methods will not be mathematically consistent and have some negative eigenvalues. A practical method for constructing the consistent correlation matrix that is closest to a forecasted one is given in Rebonato and  J\"{a}ckel (2000). For the portfolios discussed in sections 4 and 5 the $n \times n$ correlation matrix $\xi_{ij}$ is taken to have all of its off diagonal elements the same, {\it i.e.},
$\xi_{ij}=\xi$ for $i \ne j$. Such a correlation matrix has one eigenvalue equal to $1+(n-1)\xi$ and the others equal to $1-\xi$. Consequently its determinant is 
\be
{\rm det} [\xi_{ij}]=(1-\xi)^{n-1}[1+(n-1) \xi].
\ee
Its inverse has diagonal elements,
\be
\xi^{(-1)}_{ii}={1+(n-2) \xi \over 1+(n-2) \xi -(n-1)\xi^2},
\ee
and off diagonal elements ($i \ne j$),
\be
\xi^{(-1)}_{ij}=-{\xi \over 1+(n-2)\xi-(n-1)\xi^2}.
\ee

In the next section we consider the problem of portfolio allocation for portfolios consisting of corporate bonds subject to default risk and risk free assets. The implications of default risk are addressed using the model discussed in this section.
Other sources of risk, for example, systematic risk associated with the liquidity part of the excess risk premium, are neglected.

\section{The Portfolio Problem} 

Assume a zero coupon bond from company $i$ grows in value (if it doesn't default) at the (short) corporate rate $c_i(t)$ and that if the company defaults a fraction $R_i$ of the value of that bond at the time of default is reinvested at the risk free (short) rate $y(t)$. Then, the random variable for the value of this bond at some time $T$ in the future is,
\begin{eqnarray}
\label{value}
{\hat V_i(T) \over V_i(0)}&=&\exp\left(\int_0^T d\tau c_i(\tau)\right)(1-\hat n_i(T)) \nonumber  \\
&+&R_i\int_0^T ds { d \hat n_i(s) \over ds }\exp\left(\int_0^s d\tau c_i(\tau)d\tau +\int_s^T d\tau y(\tau)\right),
\end{eqnarray}
where $V_i(0)$ is the initial value of the zero coupon bond.  In equation (\ref{value}) we have continuously compounded the returns and we have assumed that $T$ is less than or equal to the maturity date of the bond\footnote{Assuming that after maturity the value of a zero coupon corporate bond is reinvested at the risk free rate it is straightforward to generalize the analysis of this paper to portfolios containing zero coupon bonds with different maturities, some of which are less than the investment horizon. Similarly default risk for portfolios containing coupon paying bonds can be studied since each coupon payment can be viewed as a zero coupon bond.}. Note that the random variable $d\hat n_i(t)/dt$ is equal to the Dirac delta function, $\delta(t-t_i)$, if company $i$ defaults at the time $t_i$ and is zero if it doesn't default. The first term in equation (\ref{value}) gives the value if the company does not default before time $T$ and the second term (proportional to $R_i$) gives the value if the company defaults before time $T$. The corresponding formula for a risk free asset is,
\be
{V_{rf}(T) \over V_{rf}(0)}=\exp \left( \int_0^T d\tau y(\tau) \right).
\ee
Note that we are not treating the (short) risk free rate $y(\tau)$ as a random variable, although it is certainly possible to generalize this formalism to do that. The (short) corporate rate is decomposed as
\be
c_i(t)=y(t)+\lambda_i(t)+\mu_i(t),
\ee
where $\mu_i$ is the excess risk premium and $\lambda_i$ is the part of corporate short rate that compensates the investor in an actuarially fair way for the fact that the value of the investment can be reduced through corporate default. In other words,
\be
\label{fairvalue}
E[\hat V_i(T)/V_i(0)]|_{\mu_i=0}=V_{rf}(T)/V_{rf}(0).
\ee
Taking the expected value equation (\ref{fairvalue}) implies that
\be
\label{nice1}
1=(1-P_i(T))\exp \left( \int_0 ^T d\tau \lambda_i(\tau)\right)+R_i\int_0^T ds {d P_i(s) \over ds}\exp \left( \int_0^s d\tau \lambda_i (\tau) \right).
\ee
Differentiation with respect to $T$ gives
\be
\label{cute}
\lambda_i(T)={dP_i(T)\over dT}{1-R_i \over 1-P_i(T)}.
\ee
Using equation (\ref{hazard}) for the time dependence of the probabilities of default equation (\ref{cute}) implies the familiar relation
\be
\lambda_i(T)=h_i(1-R_i).
\ee

The above results imply that the total wealth, after time $T$, in a portfolio consisting of risk free assets and zero coupon corporate bonds that are subject to default risk is (assuming the bonds mature at dates greater than or equal to $T$)
\begin{eqnarray}
\label{wealth}
\hat W(T)&=&W_0\exp\left(\int_0^T d\tau y(\tau)\right)\times \nonumber  \\ 
&~&\left[ (1-\alpha)+\alpha \left(\sum_i f_i\exp\left(\int_0^T d\tau(\lambda_i (\tau) + \mu_i(\tau))\right)-\hat l (T)\right) \right],
\end{eqnarray}
where, $W_0$ is the initial wealth, $\alpha$ is the fraction of corporate assets in the portfolio, $\hat l (T)$ is the random variable for the fractional default loss over the time period $T$,  $R_i$ is the recovery fraction and $f_i$ denotes the initial fraction of corporate assets in the portfolio that are in firm $i$ ($\sum_i f_i=1$). In equation (\ref{wealth}) the sum over $i$ goes over all $N$ firms in the portfolio and the default loss random variable is given by,
\begin{eqnarray}
\hat l(T)&=& \sum_i f_i \left[\exp\left(\int_0^T d\tau (\lambda_i(\tau)+\mu_i(\tau))\right)\hat n_i(T) \right. \nonumber  \\
&-&\left. R_i\int_0^T ds { d \hat n_i(s) \over ds }\exp\left(\int_0^s d\tau ((\lambda_i(\tau)+\mu_i(\tau ))\right)\right].
\end{eqnarray}
Integrating by parts,
\begin{eqnarray}
\hat l(T)&=& \sum_i f_i \left[\exp\left(\int_0^T d\tau (\lambda_i(\tau)+\mu_i(\tau))\right)\hat n_i(T)(1-R_i) \right. \nonumber  \\
&+&\left. R_i\int_0^T ds \hat n_i(s)(\lambda_i(s)+\mu_i(s))\exp\left(\int_0^s d\tau ((\lambda_i(\tau)+\mu_i(\tau ))\right)\right].
\end{eqnarray}
The expected value of $\hat l(T)$ is,
\begin{eqnarray}
\label{expvalue} 
E[\hat l(T)]&=&\sum_i f_i \left[\exp\left(\int_0^T d\tau (\lambda_i(\tau)+\mu_i(\tau))\right)P_i(T)\right. \nonumber  \\
& -&\left. R_i\int_0^T ds { d P_i(s) \over ds }\exp\left(\int_0^s d\tau ((\lambda_i(\tau)+\mu_i(\tau ))\right)\right].
\end{eqnarray}
Fluctuations of the fractional loss $\hat l (T)$ about its average value are described by the variable,
\be
\delta \hat l (T)= \hat l (T)-E[\hat l(T)].
\ee
The mean of $\delta \hat l$ is zero and the probability distribution for $\delta \hat l$ determines the default risk of the portfolio associated with fluctuations of the random variables $ \hat n_i(t)$.  The moments of this probability distribution are, 
\be
v^{(m)}(T)=E[(\delta \hat l (T))^m].
\ee
 Using equation (\ref{two}) and the property $\hat n_i(t_1) \hat n_i(t_2)=\hat n_i({\rm min}[t_1,t_2])$,
the moments $v^{(m)}(T)$ can be expressed in terms of the joint default probabilities.

The random variable for the portfolio wealth at time $T$ is,
\be
\hat W(T)=W_0 \exp\left(\int_0^T d\tau y(\tau)\right)[1+\alpha {\rm x}(T) -\alpha \delta \hat l (T)],
\ee 
where, 
\be
{\rm x}(T)=-1+\sum f_i \exp\left( \int_0^T d \tau (\lambda_i(\tau)+\mu_i(\tau))\right)-E[\hat l(T)].
\ee
 Using equations (\ref{nice1}) and (\ref{expvalue}) we find that,
\begin{eqnarray}
{\rm x(T)}&=&\sum_if_i\left[\exp\left( \int_0^T d\tau \mu_i (\tau)\right)-1  \right. \nonumber \\
&+&\left. R_i\int_0^T ds {d P_i(s) \over ds} \exp\left( \int_0^s d\tau (\lambda_i(\tau)+\mu_i(\tau))\right)\left(1-\exp \left(\int_s^T d\tau \mu_i (\tau)\right)\right)\right].
\end{eqnarray}
Note that ${\rm x}(T)$ vanishes if the excess risk premiums $\mu_i$ do. 

To find out what value of $\alpha$ is optimal for some time horizon $T$ a utility function is introduced which characterizes the investor's level of risk aversion. Here we use utility functions of the type $U{_\gamma}(W) =W^{\gamma}/\gamma$ which have constant relative risk aversion\footnote{See, for example, Ingersoll (1987).}, $1-\gamma$. The optimal fraction of corporates, $\alpha_{opt}$, maximizes the expected utility of wealth $E[U{_\gamma}(\hat W(T))]$. Expanding the utility of wealth in a power series in $\delta \hat l$ and taking the expected value gives
\begin{eqnarray}
\label{momexp}
E[U{_\gamma}(\hat W(T))]&=&(W_0^{\gamma}/\gamma)\exp\left(\gamma\int_0^Td\tau y(\tau)\right)( 1+\alpha {\rm x}(T))^{\gamma}\nonumber \\
&\times&\left[ 1+\sum_{m=2}^{\infty}{\Gamma (m-\gamma) \over \Gamma (-\gamma)\Gamma(m+1)}\left({\alpha  \over 1+\alpha {\rm x}(T)} \right)^m v^{(m)}(T) \right],
\end{eqnarray}
where $\Gamma$ is the Euler Gamma function.
The approximate optimal value of $\alpha$ obtained from truncating the sum in equation (\ref{momexp}) at the $m$'th moment is denoted by $\alpha_m$. The focus of this paper is on portfolios that are not leveraged and have $\alpha_{opt}$ less than unity. We will later see that typically, for such portfolios, the $\alpha_m$ converge very quickly to $\alpha_{opt}$ so that for practical purposes only a few of the moments $v^{(m)}$ need be calculated. Note that the value of $\alpha_{opt}$ is independent of the risk free rate $y(t)$. 

The expressions for $\hat l(T)$ and ${\rm x}(T)$ are complicated but they do simplify for short times $T$ or if all the  recovery fractions are zero. For example, if $T$ is small
\be
\hat l(T)\simeq \sum_i f_i (1-R_i)\hat n_i(T),
\ee 
and
\be
{\rm x}(T) \simeq \sum_i f_i \int_0^T ds\mu_i(s).
\ee
On the other hand if all the recovery fractions vanish  ({\it i.e.} $R_i=0$) then,
\be
\hat l(T)=\sum_i f_i \exp \left( \int_0^T d\tau (\lambda_i(\tau)+\mu_i(\tau))\right)\hat n_i(T),
\ee 
and
\be
{\rm x}(T)=\sum_i f_i \left[\exp \left( \int_0^T d\tau \mu_i(\tau)\right)-1\right].
\ee

If ${\rm x}(T)$ is zero then $\alpha_{opt}$ is also zero. It is ${\rm x}(T)$  that contains the dependence on the excess risk premiums which provide the incentive for a risk adverse investor to choose corporate bonds over risk free assets. An approximate formula for $\alpha_{opt}$ can be derived by expanding it in ${\rm x}(T)$,
\be
\label{approxalph}
\alpha_{opt} = \sum_{n=1}^{\infty}{s_n (T)\over n!} {\rm x}(T)^n.
\ee
The coefficients $s_i(T)$ can be expressed in terms of the moments, $v^{(m)}(T)$. Explicitly, for the first two coefficients,
\be
\label{ones}
s_1(T)={1\over (1-\gamma)v^{(2)}(T)},
\ee
and
\be
\label{twos}
s_2(T)=-{(2-\gamma)v^{(3)}(T)\over (1-\gamma)^2 (v ^{(2)}(T))^3}.
\ee
For $\gamma <1$ including the third moment reduces the optimal fraction of corporates when $v^{(3)}(T)$ is positive.

\section{Sample Portfolios with Zero Recovery Fractions}

Here we consider very simple sample portfolios where the correlations, default thresholds and excess risk premiums are the same for all $N$ firms in the portfolio {\it i.e.,} $\xi_{ij}=\xi$, $\chi_i(t)=\chi(t)$ and $\mu_i(t)=\mu$\footnote{We also assume that $\mu$ is independent of time.}. Then all the probabilities of default are the same, $P_i(t)=P(t)$, and the joint default probabilities are also independent of which firms are being considered, $P_{i_1 \ldots i_m}(t_1, \ldots t_m)=P_{12 \ldots m}(t_1, \ldots t_m)$. The time dependence of the default probabilities is taken to be given by equation (\ref{hazard}). Given our assumptions the hazard rates are then also independent of firm, $h_i=h$. We also take the portfolios to contain equal assets in the firms so that, $f_i=1/N$, for all $i$, and assume all the recovery fractions are zero. Since the portfolio allocation problem does not depend on the risk free rate $y$ or the initial wealth we set $y=0$ and $W_0=1$.

With these assumptions the expressions for $\hat l (T)$ and ${\rm x(T)}$ become, 
\be
\hat l(T)={1 \over N}\exp(hT+\mu T)\sum_i \hat n_i(T),
\ee
and
\be
{\rm x}(T)=\exp(\mu T)-1.
\ee

For random defaults the probability of a fractional loss of $\hat l=\exp(hT+\mu T)n/N$ in the time horizon $T$ is $(1-P(T))^{N-n}P(T)^n N!/(N-n)!n!$ and so the $m$'th moment of the default loss distribution is
\be
v^{(m)}(T)=\exp(m hT+m\mu T)\sum_{n=0}^N\left({n \over N}-P(T)\right)^m (1-P(T))^{N-n}P(T)^n {N! \over (N-n)!n!}.
\ee
The expected utility of wealth is
\begin{eqnarray}
\label{ew}
E[U_{\gamma}(\hat W(T))]&=&{1 \over \gamma}\sum_{n=0}^N (1-P(T))^{N-n}P(T)^n {N! \over (N-n)!n!} \nonumber  \\
&\times& \left(1+\alpha {\rm x}(T) -\alpha\left({n \over N}-P(T) \right)\exp(hT+\mu T)\right)^{\gamma}.
\end{eqnarray}

Having the explicit expression for the utility of wealth in equation (\ref{ew}) lets us compare results of the moment expansion for the optimal fraction of corporates $\alpha_m$ with the all orders result, $\alpha_{opt}$. These are presented in Table I in the case $h=0.02$, $\mu=100{\rm bp}$ and $\gamma=-4$. Results for different values of the number of firms $N$ and different time horizons $T$ are shown in Table I. We also give in columns three and four of Table I the
volatility
\be
{\rm vol}=\sqrt{v^{(2)}(T)},
\ee
 and skewness 
\be
{\rm skew}={v^{(3)}(T) \over \sqrt{v^{(2)}(T)}^3},
\ee
of the portfolio default loss probability distribution.

Increasing $N$ gives a larger value for the optimal fraction of corporates because diversification reduces risk. This occurs very rapidly with $N$. For $N=1$, $\mu=100 {\rm bp}$, $T=1 {\rm yr}$ and $\gamma=-4$ the optimal fraction of corporates is only $7.7\%$. By $N = 10$ increasing the number of firms has reduced the portfolio default risk so much that the $100 {\rm bp}$ excess risk premium causes a portfolio that is $76\%$ corporates to be preferred. 

For all the entries in Table I the moment expansion converges very rapidly, although for low $N$ it is the small value of $\alpha_{opt}$ that is driving the convergence. Since in equation (\ref{momexp}) the term proportional to $v^{(m)}$ has a factor of $\alpha^m$ accompanying it we expect good convergence of the moment expansion at small $\alpha$.

The focus of this paper is on portfolios that are not leveraged and have $\alpha_{opt}<1$. But a value $\alpha_{opt} > 1$ is not forbidden when finding the maximum of the expected utility of wealth. This occurs at lowest order in the moment expansion in the last row of Table I.
\vskip0.25in
\begin{center}
\noindent
Table I: Optimal Fraction of Corporates for Sample Portfolios with Random ($\xi=0$) Defaults. Other parameters used are: $h=0.02$, $\gamma=-4$ and $\mu=100 {\rm bp}$.
\vskip 0.25in
\noindent
$
\bordermatrix{& T & N& {\rm vol}&{\rm skew} &\alpha_2 & \alpha_3& \alpha_4 & \alpha_5 &\alpha_{opt}\cr
&1{\rm yr}&1 & 0.14&6.9&0.098&0.079&0.077&0.077& 0.077\cr
&1{\rm yr}&5&0.064&3.1&0.50&0.40&0.39&0.38&0.38\cr
&1{\rm yr}&10&0.045& 2.2&1.0&0.81&0.78&0.77& 0.76\cr
&5{\rm yr}&1 & 0.34&2.8&0.090&0.074& 0.072&0.072& 0.072  \cr 
&5{\rm yr}&5 &0.15&1.2&0.49&0.39&0.37&0.36&0.36\cr 
&5{\rm yr}&10& 0.11&0.87&1.1&0.85&0.75&0.73&0.71\cr
}$
\end{center}
\vskip0.25in

Next we consider the more typical case where there are correlations, $\xi \ne 0$. Table IIa gives values of $\alpha_2$ to $\alpha_5$ for $T=1{\rm yr}$, $\mu=100{\rm bp}$, $h=0.02$ and $\gamma=-4$. Portfolios with $N=10$, $50$ and $100$ are considered and values for $\xi$ between $0.5$ and $0.25$ are used.

Again the convergence of the moment expansion is quite good. Although just including the variance ({\it i.e.} $\alpha_2$) can be off by almost a factor of two $\alpha_3$ is usually a reasonable approximation to the true optimal fraction of corporates. The convergence is worse the larger the value of $\alpha_{opt}$ and for values of $\alpha_{opt}$ less than $40\%$ we find that $\alpha_3$ is within about $10\%$ of $\alpha_{opt}$.

In the last row of Table IIa the value of $\alpha_5$ is equal to unity. However, we know that the true value of
$\alpha_{opt}$ must be less than unity.  For $\alpha=1$ there is some finite (but tiny) chance of the investor loosing all his wealth and for $\gamma<0$ the utility of zero wealth is $-\infty$.

\vskip0.25in
\begin{center}
Table IIa:  Moment Expansion for Optimal Portfolio with Correlated Defaults for $T=1{\rm yr}$, $\mu=100{\rm bp}$, $h=0.02$ and $\gamma=-4$.
\vskip 0.25in
$
\bordermatrix{&T&N& \mu&\xi&d_{ij}&{\rm vol}&{\rm skew}&\alpha_2&\alpha_3&\alpha_4&\alpha_5\cr
&1{\rm yr}&10&100{\rm bp}&0.50&0.152&0.070&5.2&0.42&0.31&0.29&0.29\cr
&1{\rm yr}&10&100{\rm bp}&0.45& 0.126&0.066&4.9&0.47&0.35&0.33&0.32\cr
&1{\rm yr}&10&100{\rm bp}& 0.40&0.102&0.063&4.5&0.52&0.39&0.36&0.36 \cr
&1{\rm yr}&10&100{\rm bp}&0.35 &0.082&0.060&4.2&0.57&0.43&0.40& 0.40 \cr
&1{\rm yr}&10&100{\rm bp}&0.30&0.064&0.057&3.8&0.63&0.48&0.45&0.44\cr
&1{\rm yr}&10&100{\rm bp}&0.25&0.049&0.054&3.5&0.70 & 0.53 & 0.50 & 0.49 \cr
&1{\rm yr}&50&100{\rm bp}&0.50&0.152&0.059&5.5&0.59 & 0.42 & 0.38 & 0.37\cr
&1{\rm yr}&50&100{\rm bp}&0.45& 0.126&0.054&5.2 &0.70 & 0.49 & 0.45& 0.43\cr
&1{\rm yr}&50&100{\rm bp}& 0.40&0.102&0.050&4.8&0.84 & 0.58& 0.53 & 0.51 \cr
&1{\rm yr}&50&100{\rm bp}&0.35 &0.082&0.045&4.4&1.0 & 0.70 & 0.63& 0.61 \cr
&1{\rm yr}&50&100{\rm bp}&0.30&0.064&0.041&4.0&1.2 & 0.85 & 0.77 & 0.74 \cr
&1{\rm yr}&50&100{\rm bp}&0.25&0.049&0.037&3.6&1.5 &1.1 & 0.94 & 0.90\cr
&1{\rm yr}&100&100{\rm bp}&0.50&0.152&0.057&5.6&0.62 & 0.44 & 0.40& 0.39\cr
&1{\rm yr}&100&100{\rm bp}&0.45& 0.126&0.053&5.3& 0.75 & 0.52 & 0.47& 0.45\cr
&1{\rm yr}&100&100{\rm bp}& 0.40&0.102&0.048&5.0& 0.91 & 0.62& 0.56 & 0.54\cr
&1{\rm yr}&100&100{\rm bp}&0.35 &0.082&0.043&4.6&1.1 & 0.76 & 0.68& 0.65\cr
&1{\rm yr}&100&100{\rm bp}&0.30&0.064&0.039&4.2& 1.4 & 0.94 & 0.84 & 0.80\cr
&1{\rm yr}&100&100{\rm bp}&0.25&0.049&0.035&3.8& 1.8& 1.2 & 1.1 & 1.0\cr
}$
\end{center}
\vskip0.25in

Correlations dramatically effect the dependence of the optimal portfolio allocation on the total number of firms. For the $\xi=0.5$ entries in Table IIa the optimal fraction of corporates is, $0.29$, $0.37$, and $0.39$ for $N=10$, $50$ and $100$ respectively. For $N=10,000$ we find that $\alpha_5=0.40$. Increasing the number of firms beyond $100$ only results in a small increase in the optimal fraction of corporates. When defaults are random the moments of the portfolio default loss distribution go to zero as $N \rightarrow \infty$. For example, with $\xi=0$ the variance of the default loss distribution is,
\be
\label{vol1}
v^{(2)}=\exp(2hT+2\mu T){P(1-P) \over N},
\ee
and the skewness of the default loss distribution is
\be
\label{skew1}
{v^{(3)} \over (v^{(2)})^{3/2}}={1 \over \sqrt{NP(1-P)}}(1-2P).
\ee
For random defaults as $N \rightarrow \infty$ the distribution for $\delta \hat l$ approaches the trivial one where $\delta \hat l=0$ occurs with unit probability.
However, for $\xi >0$ the moments $v^{(m)}$ go to non-zero values in the limit $N \rightarrow \infty$ and the default loss distribution remains non-trivial and non-normal.

In Table IIb the time horizon is changed to $T=5 {\rm yr}$ but the other parameters are left the same as in Table IIa. The value of $\alpha_{opt}$ is always smaller than in Table IIa indicating that the compounding of the excess risk premium does not completely compensate for the added risk associated with the greater probability of default. The convergence of the moment expansion and the dependence of the optimal fraction of corporates on the number of bonds is similar to that in Table
IIa.

\vskip0.25in
\begin{center}
Table IIb:  Moment Expansion for Optimal Portfolio with Correlated Defaults for $T=5{\rm yr}$, $\mu=100{\rm bp}$, $h=0.02$ and $\gamma=-4$.
\vskip 0.25in
$
\bordermatrix{&T&N& \mu&\xi&d_{ij}&{\rm vol}&{\rm skew}&\alpha_2&\alpha_3&\alpha_4&\alpha_5\cr
&5{\rm yr}&10&100{\rm bp}&0.50&0.246&0.193&2.3&0.29&0.22&0.21&0.20\cr
&5{\rm yr}&10&100{\rm bp}&0.45& 0.213&0.184&2.2&0.32&0.24&0.23&0.22\cr
&5{\rm yr}&10&100{\rm bp}&0.40&0.182&0.175&2.1&0.36&0.27&0.25&0.25 \cr
&5{\rm yr}&10&100{\rm bp}&0.35 &0.153&0.166&2.0&0.40&0.30&0.28& 0.27 \cr
&5{\rm yr}&10&100{\rm bp}&0.30&0.127&0.158&1.9&0.45&0.34&0.31&0.31\cr
&5{\rm yr}&10&100{\rm bp}&0.25&0.102&0.149&1.8&0.51 & 0.38 & 0.35 & 0.34 \cr
&5{\rm yr}&50&100{\rm bp}&0.50&0.246&0.174&2.4&0.36 & 0.27 & 0.25 & 0.24\cr
&5{\rm yr}&50&100{\rm bp}&0.45& 0.213&0.163&2.3 &0.42 & 0.31 & 0.28& 0.28\cr
&5{\rm yr}&50&100{\rm bp}& 0.40&0.182&0.152&2.2&0.49 & 0.35& 0.32 & 0.31 \cr
&5{\rm yr}&50&100{\rm bp}&0.35 &0.153&0.141&2.1&0.58 & 0.41 & 0.38& 0.36 \cr
&5{\rm yr}&50&100{\rm bp}&0.30&0.127&0.129&2.0&0.71 & 0.49 & 0.44 & 0.43 \cr
&5{\rm yr}&50&100{\rm bp}&0.25&0.102&0.118&1.8&0.89 &0.60 & 0.54 & 0.51\cr
&5{\rm yr}&100&100{\rm bp}&0.50&0.246&0.172&2.4&0.38 & 0.28& 0.26& 0.25\cr
&5{\rm yr}&100&100{\rm bp}&0.45& 0.213&0.160&2.3& 0.44 & 0.32 & 0.29& 0.28\cr
&5{\rm yr}&100&100{\rm bp}& 0.40&0.182&0.149&2.2& 0.52 & 0.37& 0.33 & 0.33\cr
&5{\rm yr}&100&100{\rm bp}&0.35 &0.153&0.137&2.1&0.62 & 0.43 & 0.39& 0.38\cr
&5{\rm yr}&100&100{\rm bp}&0.30&0.127&0.125&2.0& 0.76 & 0.52 & 0.47 & 0.45\cr
&5{\rm yr}&100&100{\rm bp}&0.25&0.102&0.113&1.8& 0.98& 0.65 & 0.57 & 0.54\cr
}$
\end{center}
\vskip0.25in

Table IIb indicates that the volatility of the portfolio default loss probability distribution increases as the time horizon increases and the skewness decreases as the time horizon increases. The same behavior occurs for random defaults and this can be seem from equations (\ref{vol1}) and ({\ref{skew1}) with the small probability of default $P$ increasing with time.

Next we consider the effect of changing the utility function to one with a lower level of risk aversion. In Table III $\gamma=-2$ is used. Equation (\ref{ones}) suggests that an excess risk premium of $60 {\rm bp}$ would yield roughly the same portfolio allocation as the excess risk premium of $100 {\rm bp}$ used in Tables II. In Table III we consider portfolios with $100$ firms and take the same parameters as in Table IIa apart from the lower level of risk aversion, $\gamma=-2$ and the lower excess risk premium. $\mu=60{\rm bp}$.

It is not surprising that the values of $\alpha_2$ in Table III are close to those in Table IIa. However, $\alpha_5$ is typically about $10\%$ larger than in Table IIa. The convergence of the moment expansion is better in Table III than in
Table IIa.
\newpage
\vskip0.25in
\begin{center}
Table III:  Moment Expansion for Optimal Portfolio with Correlated Defaults for $T=1{\rm yr}$, $N=100$, $\mu=60{\rm bp}$, $h=0.02$, and $\gamma=-2$.
\vskip 0.25in
$
\bordermatrix{&T&N& \mu&\xi&d_{ij}&{\rm vol}&{\rm skew}&\alpha_2&\alpha_3&\alpha_4&\alpha_5\cr
&1{\rm yr}&100&60{\rm bp}&0.50&0.152&0.057&5.6&0.62 & 0.47 & 0.45& 0.44\cr
&1{\rm yr}&100&60{\rm bp}&0.45& 0.126&0.053&5.3& 0.74 & 0.56 & 0.53& 0.52\cr
&1{\rm yr}&100&60{\rm bp}& 0.40&0.102&0.048&5.0& 0.90 & 0.68& 0.63 & 0.62\cr
&1{\rm yr}&100&60{\rm bp}&0.35 &0.082&0.043&4.6&1.1 & 0.82 & 0.76& 0.74\cr
&1{\rm yr}&100&60{\rm bp}&0.30&0.064&0.039&4.2& 1.4 & 1.0 & 0.94 & 0.92\cr
&1{\rm yr}&100&60{\rm bp}&0.25&0.049&0.035&3.8& 1.7& 1.3 & 1.2 & 1.1\cr
}$
\end{center}
\vskip0.25in

The examples in this section have all used an annualized hazard rate of $2\%$, however, the convergence of the moment expansion is similar for significantly larger default probabilities. In Table IV the convergence of the moment expansion is examined for portfolios with $100$ corporate bonds and hazard rates of  $2\%$, $4\%$, $6\%$, $8\%$, and $10\%$. The value of $\xi$ is fixed at $0.50$ and the time horizon is $1{\rm yr}$.
The convergence of the moment expansion is good for all the values of $h$ used in Table IV. 

For longer time horizons the increased default risk is not compensated for by the compounding of the excess risk premium. For example if we take the parameters of the last row of Table IV but increase the time horizon to
$T=5 {\rm yr}$ then, $\alpha_2=0.25$, $\alpha_3=0.21$, $\alpha_4=0.18$ and $\alpha_5=0.17$.
\vskip0.25in
\begin{center}
Table IV:  Probability of Default and Convergence of the Moment Expansion for $T=1 {\rm yr}$, $\xi=0.50$, $N=100$ and $\gamma=-4$.
\vskip 0.25in
$
\bordermatrix{&T&h& \mu&\xi&d_{ij}&{\rm vol}&{\rm skew}&\alpha_2&\alpha_3&\alpha_4&\alpha_5\cr
&1{\rm yr}&0.02&100{\rm bp}&0.50&0.152&0.057&5.6&0.62 & 0.44 & 0.40& 0.39\cr
&1{\rm yr}&0.04&200{\rm bp}&0.50& 0.189&0.092&4.0& 0.50 & 0.36 & 0.33& 0.32\cr
&1{\rm yr}&0.06&300{\rm bp}& 0.50&0.214&0.121&3.2& 0.44 & 0.32& 0.29 & 0.29\cr
&1{\rm yr}&0.08&500{\rm bp}&0.50 &0.231&0.148&2.7&0.52 & 0.35 & 0.32& 0.31\cr
&1{\rm yr}&0.10&600{\rm bp}&0.50&0.246&0.173&2.4& 0.46 & 0.32 & 0.29 & 0.28\cr
}$
\end{center}
\vskip0.25in

\section{Sample Portfolios with Non-Zero Recovery Fractions}

In this section we consider portfolios like those in section 4 but allow the recovery fractions $R_i=R$ to be nonzero.  Here we focus on short time horizons using the approximate formulas,
\be
\label{rnz1}
\hat l(T)= (1-R)\hat n_i(T),
\ee 
and
\be
\label{rnz2}
{\rm x}(T) =\mu T,
\ee
appropriate to that regime. Recall that if $R$ is not zero the actuarially fair corporate spread is $\lambda=h(1-R)$. Table V shows the impact of nonzero recovery on $\alpha_{opt}$. The parameters used are $\gamma=-4$, $T= 1{\rm yr}$, $N=10$, $R=0.5$, $h=0.04$ and $\mu=50$ and $100 {\rm bp}$. Doubling the hazard rate to $0.04$ gives an actuarially fair corporate spread of $200{\rm bp}$, which is the same as in Tables II. Note that even though the actuarially fair spread in Table V is the same as in Table IIa nonzero recovery reduces the default risk and so the excess risk premium must be reduced by a factor of two to get values of $\alpha_{opt}$ close to those in Table IIa.

\vskip0.25in
\begin{center}
Table V:  Recovery and the Moment Expansion for Optimal Portfolio with Correlated Defaults using $T=1{\rm yr}$, $N=10$, $h=0.04$, $R=0.5$ and $\gamma=-4$.
\vskip 0.25in
$
\bordermatrix{&T&N& \mu&\xi&d_{ij}&{\rm vol}&{\rm skew}&\alpha_2&\alpha_3&\alpha_4&\alpha_5\cr
&1{\rm yr}&10&50{\rm bp}&0.50&0.152&0.050&4.6&0.40&0.32&0.31&0.31\cr
&1{\rm yr}&10&50{\rm bp}&0.45& 0.126&0.048&4.5&0.44&0.36&0.34&0.34\cr
&1{\rm yr}&10&50{\rm bp}& 0.40&0.102&0.046&4.3&0.49&0.39&0.38&0.38 \cr
&1{\rm yr}&10&50{\rm bp}&0.35 &0.082&0.043&4.2&0.54&0.44&0.42& 0.42 \cr
&1{\rm yr}&10&50{\rm bp}&0.30&0.064&0.041&4.0&0.60&0.48&0.47&0.46\cr
&1{\rm yr}&10&50{\rm bp}&0.25&0.049&0.039&3.8&0.67 & 0.54 & 0.52 & 0.51 \cr
&1{\rm yr}&10&100{\rm bp}&0.50&0.152&0.050&4.6&0.81 & 0.57 & 0.53 & 0.52\cr
&1{\rm yr}&10&100{\rm bp}&0.45& 0.126&0.048&4.5 &0.90 & 0.63 & 0.59& 0.57\cr
&1{\rm yr}&10&100{\rm bp}& 0.40&0.102&0.046&4.3&1.0 & 0.70& 0.65 & 0.63 \cr
&1{\rm yr}&10&100{\rm bp}&0.35 &0.082&0.043&4.2&1.1 & 0.78 & 0.72& 0.70 \cr
&1{\rm yr}&10&100{\rm bp}&0.30&0.064&0.041&4.0&1.2 & 0.86 & 0.80 & 0.78 \cr
&1{\rm yr}&10&100{\rm bp}&0.25&0.049&0.039&3.8&1.4 &0.96 & 0.88 & 0.86\cr
}$
\end{center}
\vskip0.25in

For random defaults one can easily see why recovery reduces portfolio default risk. Taking the hazard rate to be small, the time horizon to be short and the defaults to be random,
\be
v^{(2)}(T) \simeq (1-R){\lambda T \over N},
\ee
and
\be
{v^{(3)}(T) \over( v^{(2)}(T))^{3/2}}\simeq \sqrt{1-R \over N \lambda T},
\ee
where $\lambda$ is the actuarially fair spread. Hence with $\lambda$ held fixed the variance and skewness of the portfolio default loss probability distribution decrease as R increases.

Since we are using the approximate formulas in equations (\ref{rnz1}) and (\ref{rnz2}) we would not get exact agreement with Table IIa even if $R=0$. For example with $T=1{\rm yr}$, $N=10$, $\gamma=-4$, $\mu=100 {\rm bp}$, $h=0.02$, $\xi=0.5$ and $R=0$ using equations (\ref{rnz1}) and (\ref{rnz2}) yields, $\alpha_2=0.44$, $\alpha_3=0.33$, $\alpha_4=0.31$ and $\alpha_5= 0.30$. On the other hand the first row of Table IIa is, $\alpha_2=0.42$, $\alpha_3=0.31$, $\alpha_4=0.29$ and $\alpha_5= 0.29$.

\section{Concluding Remarks}

Default correlations have an important impact on portfolio default risk. Given a default model the tail of the default loss probability distribution is difficult to compute, often involving numerical simulation\footnote{See, for example, Duffie and Singleton (1999) and Das, Fong and Geng (2001).} of rare events. The first few moments of the default loss distribution are easier to calculate, and in the default model we used this computation involved some simple numerical integration. More significantly, the first few terms in the moment expansion have a familiar meaning. Investors are used to working with the classic mean, variance and skewness measures and have developed intuition for them and confidence in them. In this paper we studied the utility of the first few moments of the default loss probability distribution for the portfolio allocation problem.

The default model we use assumes a multivariate normal Copula function. However, the corporate assets themselves are not necessarily normal or lognormal and their probability distribution can have fat tails. When there are correlations the portfolio default loss distribution does not approach a trivial\footnote{The trivial distribution has $\delta \hat l=0$ occuring with unit probability.} or normal probability distribution as the number of firms in the portfolio goes to infinity. Correlations dramatically decrease the effectiveness of increasing the number of firms, $N$, in the portfolio to reduce portfolio default risk. 

We explored the convergence of the moment expansion for  the optimal fraction of corporate assets $\alpha_{opt}$.  Our work indicates that, for $\alpha_{opt}$ less than unity, the convergence of the moment expansion is quite good. The convergence of the moment expansion gets poorer as $\alpha_{opt}$ gets larger. We also examined how  $\alpha_{opt}$ depends on the level of risk aversion, the recovery fraction and investment time horizon.

The value of $\alpha_{opt}$ depends on the utility function used. In this paper we used utility functions with constant relative risk aversion, $1-\gamma$. It is possible to make other choices and while the quantitative results will be different for most practical purposes we expect that the general qualitative results should continue to hold.

Our main conclusions from the examples in Sections 4 and 5 are: 

\begin{itemize}
\item In the presence of correlated defaults, the default loss distribution is not normal. Nevertheless we find that retaining just the first few moments of the default loss probability distribution yields a good estimate of the optimal fraction of corporates. As $\alpha_{opt}$ gets smaller the convergence of the moment expansion improves. For the examples in sections 4 and 5 with $\alpha_{opt}$ near $40\%$ retaining just the mean, variance and skewness of the default loss probability distribution determined $\alpha_{opt}$ with a precision better than about $15\%$. 
\item For small numbers of corporate bonds increasing the number of firms decreases portfolio default risk. However, correlations cause this effect to saturate as the number of bonds increases and for the examples in sections 4 and 5 (with production factor or asset correlations $0.50 \ge \xi \ge 0.25$) the optimal fraction of corporates does not increase significantly if $N$ is increased beyond $100$ firms.
\item
Increasing the recovery fractions decreases portfolio risk and this is true even when the probabilities of default increase so that the actuarially fair spreads remain constant.
\item
With probabilities of default given by $P_i(T)=1-\exp(-h_iT)$ and the hazard rates $h_i$ constant, continuously compounding of a constant excess risk premium does not exactly compensate for the increase in the default probability with time. The optimal fraction of corporates decreased in going from a $T= 1{\rm yr}$ to a $T=5{\rm yr}$ investment horizon. 
\end{itemize}

The authors would like to thank their colleagues at PIMCO for enlightening discussions. MBW was supported in part by the Department of Energy under contract DE-FG03-92-ER40701.

\vskip0.25in

\noindent
{\Large{\bf References}}
\vskip0.25in

\noindent
Altman, E.I., (1989) {\it Measuring Corporate Bond Mortality and Performance}, Journal of Finance, Vol. 44, No. 4, 909-921.

\vspace{0.2cm}

\noindent
Black, F., and Cox, J. (1976) {\it Valuing Corporate Securities: Some Effects of Bond Indenture Provisions}, Journal of Finance, 31, 351-367.

\vspace{0.2cm}

\noindent
Das, S., Fong, G. and Geng, G. (2001) {\it The Impact of Correlated Default Risk on Credit Portfolios}, working paper, Department of Finance Santa Clara University and Gifford Fong Associates.

\vspace{0.2cm}

\noindent
Das, S., Freed, L., Geng, G. and Kapadia, N. (2001) {\it Correlated Default Risk}, working paper, Department of Finance Santa Clara University and Gifford Fong Associates.

\vspace{0.2cm}

\noindent
Das, S., and Geng, G. (2002) {\it Modeling The Process of Correlated Default}, working paper, Department of Finance Santa Clara University and Gifford Fong Associates.

\vspace{0.2cm}

\noindent
Duffie, D. and Singleton, K. (1999) {\it Simulating Correlated Defaults}, working paper, Stanford University Graduate School of Business.

\vspace{0.2cm}

\noindent
Dynkin, L., Hyman, J. and Konstantinovsky, V. (2002) {\it Sufficient Diversification in Credit Portfolios}, Lehman Brothers Fixed Income Research.

\vspace{0.2cm}

\noindent
Ingersoll, J. (1987) {\it Theory of Financial Decision Making}, Rowman and Littlefield Publishers Inc.

\vspace{0.2cm}

\noindent
Jorion, P. (2001) {\it Value at Risk}, Second Edition, McGraw-Hill Inc.

\vspace{0.2cm}

\noindent
Li, D. (2000) {\it On Default Correlation: A Copula Function Approach}, Working Paper 99-07, The Risk Metrics Group.

\vspace{0.2cm}

\noindent
Leland, H. and Toft, K. (1996) {\it Optimal Capital Structure, Endogenous Bankruptcy and the Term Structure of Credit Spreads}, Journal of Finance, 51, 987-1019.

\vspace{0.2cm}

\noindent
Longstaff, F. and Schwartz, E. (1995) {\it A Simple Approach to Valuing Risky Floating Rate Debt}, Journal of Finance, 50, 789-819.

\vspace{0.2cm}
\noindent
Lucas, D. (1995), {\it Default Correlation and Credit Analysis}, Journal of Fixed Income, March, 76-87.

\vspace{0.2cm}

\noindent
Merton, R. (1974), {\it On Pricing of Corporate Debt: The Risk Structure of Interest Rates}, Journal of Finance 29, 449-470.

\vspace{0.2cm}

\noindent
Rebonato, R. and  J\"{a}ckel, P. (2000) {\it The Most General Methodology for Creating a Valid Correlation Matrix for Risk Management and Option Pricing Purposes}, The Journal of Risk, Vol. 2, No. 2, 17-26.

\vspace{0.2cm}

\noindent
Wise, M. and Bhansali, V. (2002) {\it Portfolio Allocation To Corporate Bonds with Correlated Defaults}, To appear in the Journal of Risk.

\vspace{0.2cm}

\noindent
Zhou, C. (2001) {\it An Analysis of Default Correlations and Multiple Defaults}, The Review of Financial Studies, Vol. 12, No. 2, 555-576.

\vspace{0.2cm}
\end{document}